\documentclass[preprints,article,accept,oneauthor,pdftex]{Definitions/mdpi} 

\firstpage{1}
\usepackage{graphicx}
\usepackage{bm,amsfonts,amsthm,placeins}
\newcommand{\gl}[1]{(\ref{#1})}

\newtheorem{thm}{Condition}
\pubvolume{xx}
\articlenumber{5}
\pubyear{2019}
\copyrightyear{2019}
\history{Received: date; Accepted: date; Published: date}

\Title{Structural Distortion Stabilizing the Antiferromagnetic
  and Semiconducting Ground State of NiO}

\Author{Ekkehard Kr\"uger \orcidA{}}

\AuthorNames{Firstname Lastname, Firstname Lastname and Firstname Lastname}

\address[1]{%
Institut f{\"u}r Materialwissenschaft, Materialphysik,
  Universit{\"a}t Stuttgart, D-70569 Stuttgart, Germany}

\corres{Correspondence: ekkehard.krueger@imw.uni-stuttgart.de}

\abstract{We report evidence that the experimentally observed small
  deformation of antiferromagnetic NiO modifies the symmetry of the
  crystal in such a way that the antiferromagnetic state becomes an
  {\em eigenstate} of the electronic Hamiltonian. This deformation
  closely resembles a rhombohedral contraction, but does not possess
  the perfect symmetry of a trigonal (rhombohedral) space group. We
  determine the monoclinic base-centered magnetic space group of the
  antiferromagnetic structure within the deformed crystal which is
  strongly influenced by the time-inversion symmetry of the
  Hamiltonian. The antiferromagnetic state is evidently stabilized by
  a nonadiabatic atomic-like motion of the electrons near the Fermi
  level. This atomic-like motion is characterized by the symmetry of
  the Bloch functions near the Fermi level and provides in NiO a
  perfect basis for a Mott insulator.  }

\begin{document}


\section{Introduction}
Nickel monoxide is antiferromagnetic with the relatively high N\'eel
temperature $T_N = 523K$. Above $T_N$, NiO possesses the fcc structure
$Fm3m = \Gamma^f_cO^5_h$ bearing the international number
225~\cite{rooksby}. Cracknell and Joshua~\cite{cj} found that, below
$T_N$, the magnetic structure is invariant under the magnetic group
$C_c2/c$ which will be given explicitly in Eq.~\gl{eq:7}. The
antiferromagnetic state is accompanied by a small contraction of the
crystal along one of the triad axes often referred to as a
rhombohedral deformation. However, the magnetic group $C_c2/c$
does not possess any trigonal (rhombohedral) subgroup. Thus, this
interpretation, if taken literally, seems to imply that the ground
state of NiO does not possess any symmetry because, clearly, it cannot
have two space groups.

In Sec.~\ref{sec:rhombohedral}, this paper treats a new path in
interpreting the experimental observation of Rooksby~\cite{rooksby}:
the time-inversion symmetry of the electron system suggests that the
crystal is deformed by a small contraction closely {\em resembling}
a rhombohedral deformation of the oxygen atoms in such a way that both
the magnetic structure and the rhombohedral-like deformation have a
common magnetic space group, namely the group $M_9$ which will be
defined in Eq.~\gl{eq:15}.  Now, having determined explicitly the
magnetic group of the ground state of NiO, the group-theoretical
nonadiabatic Heisenberg model (NHM)~\cite{enhm} becomes applicable.
The NHM defines a nonadiabatic atomic-like motion shortly described in
Sec.~\ref{sec:atomiclike}. On the basis of the symmetry of the Bloch
functions in the band structure of NiO, the NHM predicts that, indeed,
a magnetic structure with the magnetic group $M_9$ may be stable in
NiO (Sec.~\ref{sec:mband}).

NiO has a second very interesting feature: it is an antiferromagnetic
Mott insulator~\cite{gavriliuk,mott_1949}. Also this observation may
be understood within the NHM. In Sec.~\ref{sec:mottinsulator} we will
show that the atomic-like motion of the electrons in antiferromagnetic
NiO stabilizes not only the antiferromagnetic state but, in addition,
provides an ideal precondition for the Mott condition to be effective
in NiO.

In order to understand the interesting features of NiO, the paper
formulates three conditions of stability. The first two conditions in
Secs.~\ref{sec:first_stab_cond} and~\ref{sec:atomiclike},
respectively, concerning the stability of a magnetic state, have
already been published in former papers. They are reformulated in
order to facilitate the reading of the paper. The third condition in
Sec.~\ref{sec:mottinsulator} concerning the existence of a Mott
insulator, is given for the first time in this paper.

\section{Group-theoretical and computational methods used in the
  paper}
\label{sec:grouptheory}
The band structure of paramagnetic NiO in Fig.~\ref{fig:bandstr225} is
calculated by the FHI-aims (``Fritz Haber Institute ab initio
molecular simulations'') computer program using the density functional
theory (DFT)~\cite{blum1,blum2} to compute the total energy in the
electronic ground state. The FHI-aims program uses spherical harmonics
as eigenvectors and provides the possibility of an output of the
eigenvectors at any wave vector $\bm k$. Thus, I was able to write a
C++ program to determine the symmetry of the Bloch functions at the
points of symmetry in the Brillouin zone for the fcc Bravais lattice
$\Gamma^f_c$ using the symmetry of the spherical harmonics as given in
Ref.~\cite{bc}. The results are given in Fig.~\ref{fig:bandstr225}.

In Table~\ref{tab:wf_9} I give the symmetry of those Bloch functions
which can be unitarily transformed into optimally localized Wannier
functions adapted to the symmetry of the magnetic group $M_9$ of
antiferromagnetic NiO. The determination of these ``magnetic bands''
(Sec.~\ref{sec:mband}) is a complex process described in
Ref.~\cite{theoriewf} (see Theorems 5 and 7 {\em ibidem}) which is
also performed by a C++ computer program.  The notes to
Table~\ref{tab:wf_9} give short remarks on the (symmetry) properties
of the Wannier functions.

\section{Magnetic group of the antiferromagnetic state - first
  stability condition}
\label{sec:first_stab_cond}
The antiferromagnetic structure of NiO is invariant under the symmetry
operations of the type IV Shubnikov space group $C_c2/c$~\cite{cj}
which may be written as~\cite{bc}
\begin{equation}
  \label{eq:7}
  C_c2/c = C2/c + K\{E|\bm \tau \}C2/c,
\end{equation}
where $K$ denotes the anti-unitary operator of time-inversion.  The
unitary subgroup $C2/c$ has the monoclinic base-centered Bravais
lattice $\Gamma^b_m$ and contains (besides the pure translations) 4
elements,
\begin{equation}
  \label{eq:9}
C2/c = \Big\{\{E|\bm 0\}, \{C_{2b}|\bm \tau\}, \{I|\bm 0\},
\{\sigma_{db}|\bm \tau\}\Big\},   
\end{equation}
when the magnetic structure is orientated as given in
Fig.\ref{fig:structures} (and in Ref.~\cite{cj}).

As in our former papers, we write the symmetry operations
$\{R|\bm t\}$ in the Seitz notation: $R$ is a point group operation
and $\bm t$ the subsequent translation. In this paper, $R$ stands for
the identity $E$, the inversion $I$, the rotation $C_{2b}$ through
$\pi$ indicated in Fig.~\ref{fig:structures}, or the reflection
$\sigma_{db} = IC_{2b}$; the translation is $\bm t = \bm 0$ or
$\bm t = \bm \tau$, where
\begin{equation}
  \label{eq:8}
  {\bm \tau} = \frac{1}{2}{\bm T}_1
\end{equation}
stands for the non-primitive translation in the group $C2/c$ indicated in
Fig.~\ref{fig:structures}.
In what follows the magnetic group $C_c2/c$ is referred to as $M_{15}$
because the unitary subgroup $C2/c$ bears the international number 15,
\begin{equation}
  \label{eq:31}
  M_{15} = C2/c + K\{E|\bm \tau \}C2/c.  
\end{equation}
Though the magnetic group $M_{15}$ leaves invariant the
antiferromagnetic structure of NiO, it need not be the the magnetic
group of antiferromagnetic NiO. This statement can be understood as
follows: Consider a magnetic material and let be
\begin{equation}
  \label{eq:10}
M = S + K\{R|\bm t\}S  
\end{equation}
a magnetic group leaving invariant the magnetic structure in this
material. $S$ is the unitary subgroup of $M$, $K$ is still the
operator of time inversion and $R$ a point group operation. $M$
includes all magnetic groups whether they are of type II, III, or
IV~\cite{bc}. Further let be $|G\rangle$ the {\em exact} magnetic
ground state of the electronic Hamiltonian $\mathcal H$. Since
$\mathcal H$ commutes with the symmetry operators $P(a)$ assigned to
the symmetry operations $a$ of $M$,
\begin{equation}
  \label{eq:11}
  [\mathcal H, P(a)] = 0\ \ \text{ for }\ \ a \in M,
\end{equation}
the magnetic state $|G\rangle$ is basis function of a one-dimensional
co-representation $D$ of $M$,
\begin{equation}
  \label{eq:12}
  P(a)|G\rangle = c(a)|G\rangle\ \ \text{ for }\ \ a \in M, 
\end{equation}
where $|c(a)| = 1$. The operators $P(a)$ are defined in
Refs.~\cite{enhm} and~\cite{theoriewf}, in the present paper there
definition is omitted.

The time-inverted state $K|G\rangle$ represents the opposite magnetic
structure and, hence, is different from $|G\rangle$,
\begin{equation}
  \label{eq:13}
  K|G\rangle \neq |G\rangle. 
\end{equation}
$K|G\rangle$ is also an eigenstate of $\mathcal H$ since $\mathcal H$
commutes with $K$,
\begin{equation}
  \label{eq:14}
  [\mathcal H, K] = 0.
\end{equation}
Hence, the states $|G\rangle$ and $K|G\rangle$ form a basis of a
two-dimensional co-representation $\widetilde D$ of the overgroup
\begin{equation}
  \label{eq:20}
  \widetilde M = M + KM
\end{equation}
of $M$, where $D$ is subduced from $\widetilde D$. Now we can
formulate a stability condition for magnetic states: the states
$|G\rangle$ and $K|G\rangle$ are {\em eigenstates} of $\mathcal H$
(i.e., $|G\rangle$ and $K|G\rangle$ represent {\em stable} magnetic
structures) if and only if the two-dimensional co-representation
$\widetilde D$ is {\em irreducible}~\cite{ea}. This statement is
well-known in the theory of ordinary (unitary) groups~\cite{wigner}.

Fortunately, it is not very complicated to decide whether or not the
magnetic group $\widetilde M$ has at least one co-representations
$\widetilde D$ complying with these conditions~\cite{bafe2as2}:
\begin{thm}
\nonumber
\label{theorem}
The group $M$ in Eq.~\gl{eq:10} may be the magnetic group of a {\em
  stable} magnetic structure if the unitary subgroup $S$ has at least
one one-dimensional single-valued representation
\begin{enumerate} 
\item 
following case (a) with respect to the magnetic group 
$S + K\{R|\bm t\}S$ in Eq.~\gl{eq:10} and
\item 
following case (c) with
respect to the magnetic group $S + KS$.
\end{enumerate}
\end{thm}
The cases (a) and (c) are defined by Eqs.\ (7.3.45) and (7.3.47),
respectively, of Ref.~\cite{bc}.

Tables~\ref{tab:rep_15} and~\ref{tab:rep_9} provide all the
information we now need for antiferromagnetic NiO: first,
Table~\ref{tab:rep_15} shows that the space group $C2/c$ (15) has only
{\em real} one-dimensional representations, and, hence, no
representation meets the second condition (ii). Consequently, a spin
structure with the magnetic group $M_{15}$ cannot be stable in NiO.

Removing from $C2/c$ the two symmetry operations $\{C_{2b}|\bm \tau\}$
and $\{I|\bm 0\}$, we receive the space group $Cc$ (9) containing
(besides the pure translations) 2 elements,
\begin{equation}
  \label{eq:23}
Cc = \Big\{\{E|\bm 0\}, \{\sigma_{db}|\bm \tau\}\Big\}.   
\end{equation}
Table~\ref{tab:rep_9} shows that the representations at points $A$ and
$M$ in the Brillouin zone of $Cc$ (9) meet condition (ii). In addition,
the first condition (i) is satisfied for the magnetic group
\begin{equation}
  \label{eq:15}
  M_9 = Cc + K\{C_{2b}|\bm 0\}Cc
\end{equation}
while it is not satisfied for $Cc + K\{E|\bm \tau\}Cc$. Consequently,
{\em the group $M_9$ is the only magnetic group in antiferromagnetic
  NiO representing a stable antiferromagnetic structure.}  Just as
$M_{15}$, the group $M_9$ has the monoclinic base-centered Bravais
lattice $\Gamma^b_m$~\cite{bc}.

Magnetoscriction alone produces the magnetic group
$M_{15}$ in NiO. Consequently, in addition to the magnetoscriction,
the crystal must be distorted in such a way that the Hamiltonian
$\mathcal H$ still commutes with the elements of
\begin{equation}
  \label{eq:16}
M_9 = \Big\{\{E|\bm 0\}, \{\sigma_{db}|\bm \tau\}, K\{C_{2b}|\bm 0\},
K\{I|\bm \tau\}, n_1{\bm T_1} + n_2{\bm T_2} + n_3{\bm T_3}\Big\},   
\end{equation}
\begin{equation}
  \label{eq:18}
[\mathcal H, P(a)]  =  0\ \ \text{for}\ \ a \in M_9,
\end{equation}
but does not commute with the symmetry operations of
\begin{equation}
  \label{eq:17}
  M_{15} -  M_9 = \Big\{\{C_{2b}|\bm \tau\}, \{I|\bm 0\}, K\{E|\bm \tau\}, K\{\sigma_{db}|\bm 0\}\Big\},
\end{equation}
\begin{equation}
  \label{eq:19}
  [\mathcal H, P(a)]  \neq  0\ \ \text{for}\ \ a \in M_{15} - M_9.
\end{equation}
This is achieved by exactly the one distortion of the crystal
illustrated in Fig.~\ref{fig:structures}: The Ni atoms are shifted in
$\pm (\bm T_2 - \bm T_3)$ direction from their positions at the lattice
points $\bm t_{Ni}$ in Eq.~\gl{eq:1}, realizing in this way the
group $M_9$ in the sense that the two commutator relations~\gl{eq:18}
and~\gl{eq:19} are satisfied.  With our group-theoretical methods we
cannot determine the magnitude of the displacements, however, they are
clearly not so large as plotted (for the sake of clarity) in
Fig.~\ref{fig:structures} (a).  The oxygen atoms, on the other hand,
are {\em not} shifted from their positions at the lattice points
$\bm t_{O}$ in Eq.~\gl{eq:4} since any dislocation of the oxygen
atoms would destroy the symmetry of the group $M_9$. These statements
on the atomic positions in the group $M_9$ may be understood by
inspection of Fig.~\ref{fig:structures}. However, they may also be
justified in terms of the Wyckoff positions of Ni and O in the group
$M_9$, see Appendix~\ref{sec:wyckoff}.

\section{Rhombohedral-like distortion}
\label{sec:rhombohedral}
\begin{figure}[!]
\centering
\includegraphics[width=.8\textwidth]{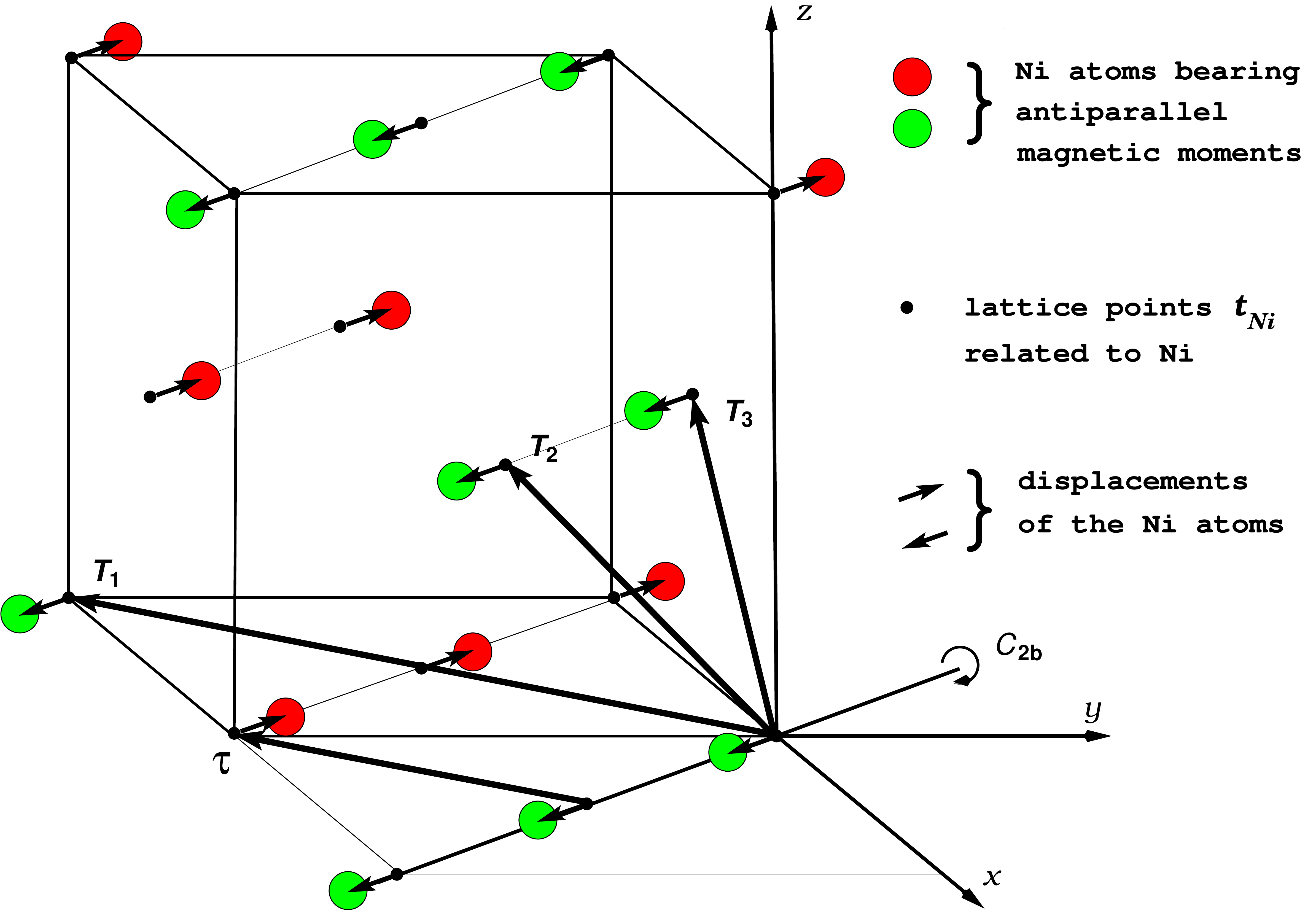}
\begin{center} (a) \end{center}
\includegraphics[width=.8\textwidth]{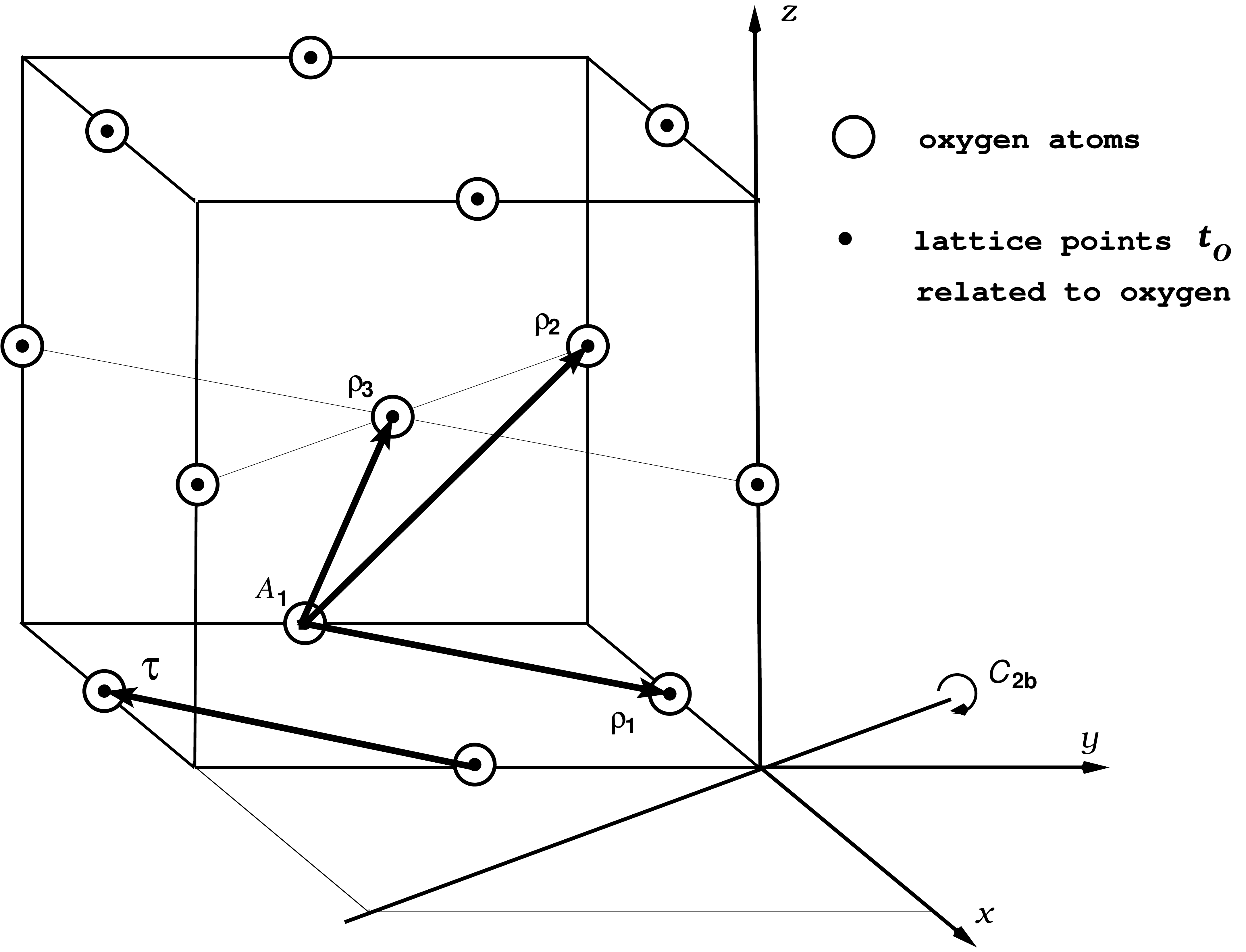}
\begin{center} (b) \end{center}
\caption{Nickel (a) and oxygen (b) atoms in distorted
  antiferromagnetic NiO with the magnetic group 
  $M_9$ in Eq.~\gl{eq:15} possessing the monoclinic base-centered
  Bravais lattice $\Gamma^b_m$. The Ni atoms
  represented by red circles bear a magnetic moment parallel or
  antiparallel to $[11\overline{2}]$ and the atoms represented by
  green circles the opposite moment. The magnetic structure is
  orientated as in Ref.~\cite{cj}. The vectors $\bm T_i$ are
  the basic translations of $\Gamma^b_m$. 
\label{fig:structures}
}
\end{figure}

Antiferromagnetic NiO becomes slightly deformed by a small contraction
along one of the triad axes~\cite{rooksby}. This deformation is often
referred to as distortion from the cubic structure in the paramagnetic
state to a rhombohedral one in the antiferromagnetic
state~\cite{slack}. On the basis of Fig.~\ref{fig:structures}, this
important and interesting experimental observation can be understood
as follows.

Fig.~\ref{fig:structures} shows exhaustively the distorted
antiferromagnetic structure of NiO with the magnetic group $M_9$. But
it should be noted that, for the sake of clarity, the basic vectors of
the Bravais lattice $\Gamma^b_m$ of $M_9$ are embedded in the
paramagnetic fcc lattice of NiO. However, the fcc lattice may be
distorted as a whole on condition that the vectors ${\bm T_1}$,
${\bm T_2}$, and ${\bm T_3}$ stay basic vectors of $\Gamma^b_m$.  The
lattice points ${\bm t}_{Ni}$ and ${\bm t}_{O}$ plotted in
Figs.~\ref{fig:structures} (a) and (b), respectively, are no longer
the positions of Ni and O in the fcc lattice, but are defined by the
equations
\begin{eqnarray}
  \label{eq:1}
  {\bm t}_{Ni} & = & n_1{\bm T_1} + n_2{\bm T_2} + n_3{\bm T_3}\ \ \ \text{ and }\nonumber\\
  {\bm t}_{Ni} & = & \frac{1}{2}\bm T_1 + n_1{\bm T_1} + n_2{\bm T_2} + n_3{\bm T_3},
\end{eqnarray}
and
\begin{eqnarray}
  \label{eq:4}
  {\bm t}_{O} & = & \frac{1}{2}({\bm T_2} - {\bm T_3} + \frac{1}{2}{\bm
  T_1}) + n_1{\bm T_1} + n_2{\bm T_2} + n_3{\bm T_3}\ \ \ \text{ and }\nonumber\\
  {\bm t}_{O} & = & \frac{1}{2}({\bm T_2} - {\bm T_3} + \frac{1}{2}{\bm
  T_1}) + \frac{1}{2}\bm T_1 + n_1{\bm T_1} + n_2{\bm T_2} + n_3{\bm T_3},
\end{eqnarray}
where ${\bm T_1}$, ${\bm T_2}$, and ${\bm T_3}$ are the basic vectors
of $\Gamma^b_m$, and $n_1$, $n_2$, and $n_3$ are integers.  Thus, the
vectors ${\bm t}_{Ni}$ and ${\bm t}_{O}$ are solely given in terms of
the basic vectors of $\Gamma^b_m$ detached from the paramagnetic fcc
lattice.

In the stable antiferromagnetic structure the Ni atoms are shifted in
$\pm$ (${\bm T_2} - {\bm T_3}$) direction from their positions at the
lattice points ${\bm t}_{Ni}$, while the oxygen atoms stay on the
positions ${\bm t}_{O}$ (Sec.~\ref{sec:first_stab_cond}).  Within a
ferromagnetic sheet, all the Ni atoms are dislocated in the same
direction. Hence, the atomic distances within a sheet are the same as
in the paramagnetic fcc phase. In adjacent sheets, on the other hand,
the dislocations have different directions and, consequently, the
distances between the Ni atoms in adjacent sheets becomes slightly
greater than in the paramagnetic phase. Hence, the initial fcc
structure is mostly disturbed in <111> direction. It is conceivable
that the antiferromagnetic structure is slightly contracted along the
<111> axis because no symmetry operation of $M_9$ forbids such a
contraction.

Fig.~\ref{fig:structures} (b) shows the position vectors
${\bm \rho}_1$, ${\bm \rho}_2$, and ${\bm \rho}_3$ of three oxygen
atoms in relation to the atom at position $A_1$. In the paramagnetic
fcc phase, the ${\bm \rho}_i$ form a basis of the trigonal
(rhombohedral) lattice $\Gamma_{rh}$ orientated in <111> direction.
Within the Bravais lattice $\Gamma^b_m$ of $M_9$, however, they are no
longer translational symmetry operations. Thus, in the
antiferromagnetc state of NiO, the ${\bm \rho}_i$ no longer define a
trigonal space group but only define the {\em positions} of the oxygen
atoms in the lattice $\Gamma^b_m$. A (slight) contraction of the
crystal along <111> has the consequence that the three position
vectors ${\bm \rho}_i$ are modified. However, they can only be
modified in such a way that the magnetic group $M_9$ is preserved,
that means that the vectors ${\bm T_1}$, ${\bm T_2}$, and ${\bm T_3}$
stay basic vectors of $\Gamma^b_m$.  Thus, the directions and the
lengths of the vectors ${\bm T_1}$, ${\bm T_2}$, and ${\bm T_3}$ are
modified on condition that
\begin{itemize}
\item ${\bm T_1}$ still passes through the plain $(\overline{1}10)$ and,
\item in relation of this plane, ${\bm T_2}$ and ${\bm T_3}$ stay
  symmetrical to one another.
\end{itemize}

The ${\bm T_i}$ and the ${\bm \rho}_i$ are connected by
the equations
\begin{eqnarray}
  {\bm T}_1 & = & -2{\bm \rho}_1\label{eq:2}\\
  {\bm T}_2 & = & {\bm \rho}_3 - {\bm \rho}_1\label{eq:3}\\
  {\bm T}_3 & = & {\bm \rho}_2 - {\bm \rho}_1\label{eq:5},
\end{eqnarray}
see Fig.~\ref{fig:structures}. From the last the two equations it
follows that
\begin{equation}
  \label{eq:6}
  {\bm T}_2 - {\bm T}_3 = {\bm \rho}_3 - {\bm \rho}_2.
\end{equation}
Let be $P$ the plane parallel to $(\overline{1}10)$ containing the
point $A_1$ in~Fig.~\ref{fig:structures} (b). Eqs.~\gl{eq:2} and
~\gl{eq:6} immediately show that ${\bm T_1}$, ${\bm T_2}$, and
${\bm T_3}$ still are basic vectors of $\Gamma^b_m$ in the disturbed
system if
\begin{enumerate}
  \item ${\bm \rho}_1$ passes through $P$ and
  \item  ${\bm \rho}_2$ and ${\bm \rho}_3$ are symmetrical to each
    other with respect to $P$.
\end{enumerate}
These two conditions are satisfied when still the modified vectors
${\bm \rho_i}$ comply with a trigonal basis orientated in
${\bm \rho}_1 + {\bm \rho}_2 + {\bm \rho}_3$ direction. Hence, the
magnetic group $M_9$ with the monoclinic base-centered Bravais lattice
$\Gamma^b_m$ is preserved when the modified ${\bm \rho}_i$ define a
rhombohedral-like array of the oxygen atoms within the antiferromagnetic state.
The rhombohedral-like array of the oxygen atoms forms an ``inner''
deformation of the oxygen atoms within the monoclinic base-centered
Bravais lattice $\Gamma^b_m$.

In summary: in the forgoing Sec.~\ref{sec:first_stab_cond} we reported
evidence that the antiferromagnetic ground state of NiO is stable only
if $M_9$ in Eq.~\gl{eq:15} is the magnetic group of the
antiferromagnetic structure. Thus, the experimentally observed
``rhombohedral structure'' is evidently the described rhombohedral-like
contraction of the crystal in
${\bm \rho}_1 + {\bm \rho}_2 + {\bm \rho}_3$ direction producing a
rhombohedral-like array of the oxygen atoms preserving the magnetic
group $M_9$ with the monoclinic base-centered Bravais lattice
$\Gamma^b_m$. Since this contraction is small, the
${\bm \rho}_1 + {\bm \rho}_2 + {\bm \rho}_3$ direction only differs
slightly from the <111> direction.

The above conditions (i) and (ii) do not require that the vectors
${\bm \rho}_i$ comply {\em exactly} with a trigonal basis. Thus, the
oxygen atoms will not form an exact trigonal array within the
antiferromagnetic system because there is no symmetry operation in
$M_9$ requiring such an exact array. Nevertheless, the conditions (i)
and (ii) allow only small deviations from an ideal trigonal array in
the antiferromagnetic system, in particular also since the vectors
${\bm \rho}_i$ form an exact trigonal basis in the paramagnetic phase
of the crystal. 


\section{Wannier functions symmetry-adapted to {\em M$_{9}$} - second
  stability condition}
\label{sec:wf}
\subsection{Atomic-like motion}
\label{sec:atomiclike}
The nonadiabatic Heisenberg model (NHM)~\cite{enhm} is based on three
immediately obvious postulates defining a strongly correlated
atomic-like motion~\cite{mott,hubbard} in narrow, partly filled bands
which cannot be described within the adiabatic approximation. The
nonadiabatic atomic-like motion is clearly separated from any
adiabatic band-like motion because the electrons gain the nonadiabatic
condensation energy $\Delta E$ (Eq.~(2.20) of Ref.~\cite{enhm}) at the
transition from the adiabatic band-like to the nonadiabatic
atomic-like motion.

The nonadiabatic localized states belonging to the atomic-like motion
are represented by hypothetical nonadiabatic localized functions. They
are adapted to the symmetry of the crystal in order that the
nonadiabatic Hamiltonian of the atomic-like system has the correct
symmetry, i.e., the correct commutation properties~\cite{enhm}. Their
existence, their spin dependence, and their symmetry is fixed by the
postulates of the NHM. So they have the same symmetry and the same
spin dependence as the {\em exact} symmetry-adapted optimally
localized Wannier functions related to one of the narrowest, partly
filled bands of the considered metal. The adjective ``exact'' means
that the Wannier functions are an exact unitary transformation of the
actual Bloch functions of the considered band. Particularly, any
modification of the symmetry of the Bloch functions in order to obtain
closed bands is not allowed because we would lose important physical
information. The known complication that the narrowest bands of the
metals are generally not closed may be solved in specific cases by
allowing the Wannier functions to have a reduced
symmetry~\cite{theoriewf}. So far, we found narrow, partly filled
bands with optimally localized symmetry-adapted Wannier functions in
the band structures of magnetic materials and of superconductors by
allowing the Wannier functions
\begin{enumerate}
\item to be adapted only to the magnetic group $M$ of the magnetic
  structure or
\item to be spin dependent,
\end{enumerate}
respectively.  An energy band with Wannier functions of the first type
(i) and the second type (ii) we called ``magnetic band related the the
magnetic group $M$'' and ``superconducting band'', respectively
(Definitions 16 and 22 of Ref.~\cite{theoriewf}), because the strongly
correlated nonadiabatic atomic-like motion in a magnetic band and in a
superconducting band evidently stabilizes
magnetism~\cite{ea,ef,bamn2as2} and
superconductivity~\cite{josn,kspace}, respectively. While
superconducting bands are not subject of this paper, the meaning of
magnetic bands shall be clarified as follows:
\begin{thm}
  A magnetic structure with the magnetic group $M$ may be stable in a
  material if and only if there exists a narrow, roughly half-filled
  magnetic band in the band structure of this material. The magnetic
  band is related to $M$ and the Wannier functions are centered at the
  positions of the atoms bearing the magnetic moments.
\end{thm}

\subsection{Magnetic band of NiO}
\label{sec:mband}


 \begin{figure*}[!]
 \includegraphics[width=.85\textwidth,angle=0]{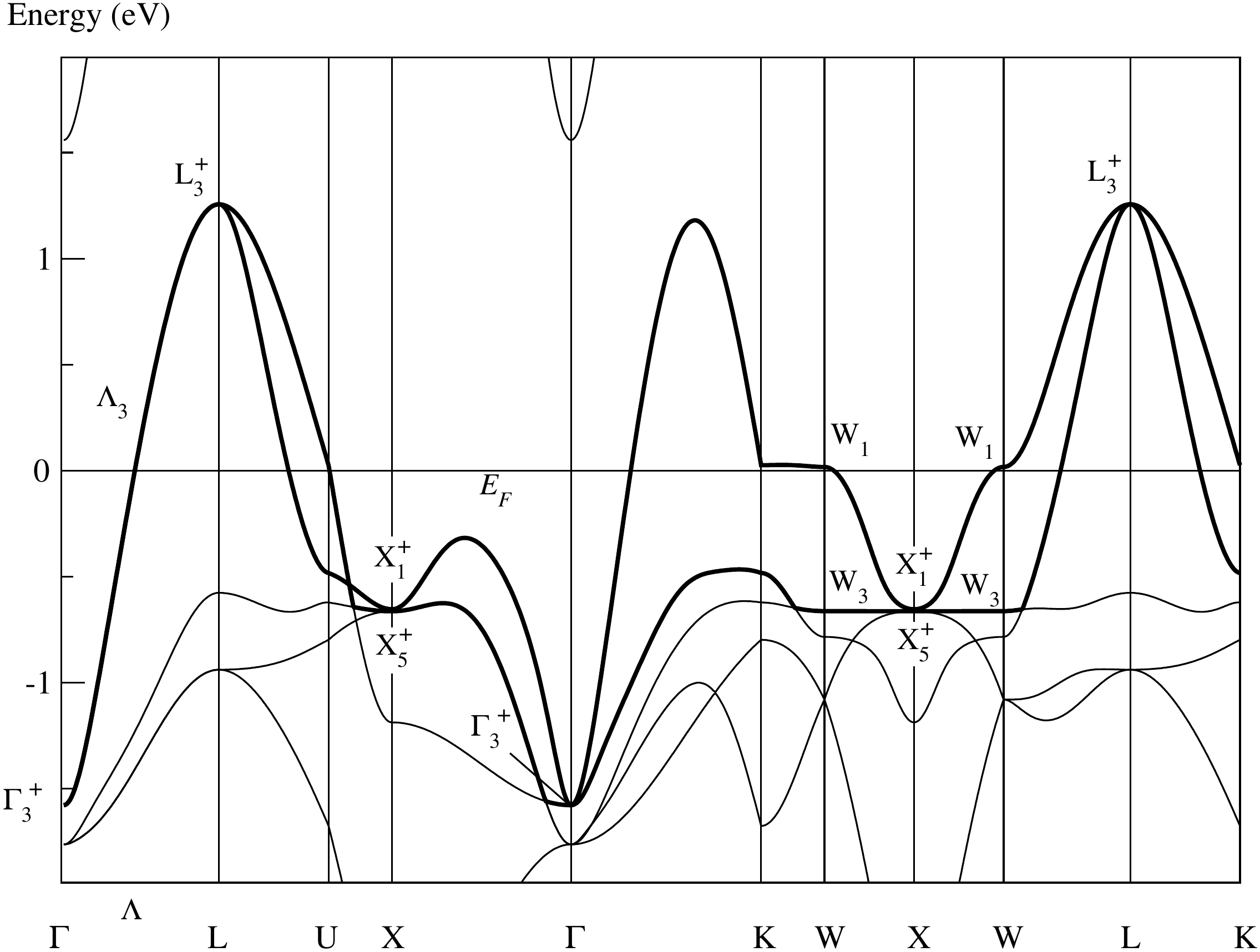}%
 \caption{ Band structure of paramagnetic fcc NiO as calculated by the FHI-aims program
\cite{blum1,blum2}, using the structure parameters given in
Ref.~\cite{rooksby}, with symmetry labels determined by the
author. The notations of the points of symmetry in the Brillouin zone
for $\Gamma^f_c$ follow Fig. 3.14 of Ref.~\cite{bc} and the symmetry
labels are defined in
Table~\ref{tab:rep_225}. The ``active'' band highlighted by the bold line becomes
a magnetic super band when folded into the Brillouin zone for the magnetic
structure, see Fig.~\ref{fig:bandstr9}.
 \label{fig:bandstr225}
}
 \end{figure*}

 
 \begin{figure*}[!]
 \includegraphics[width=.95\textwidth,angle=0]{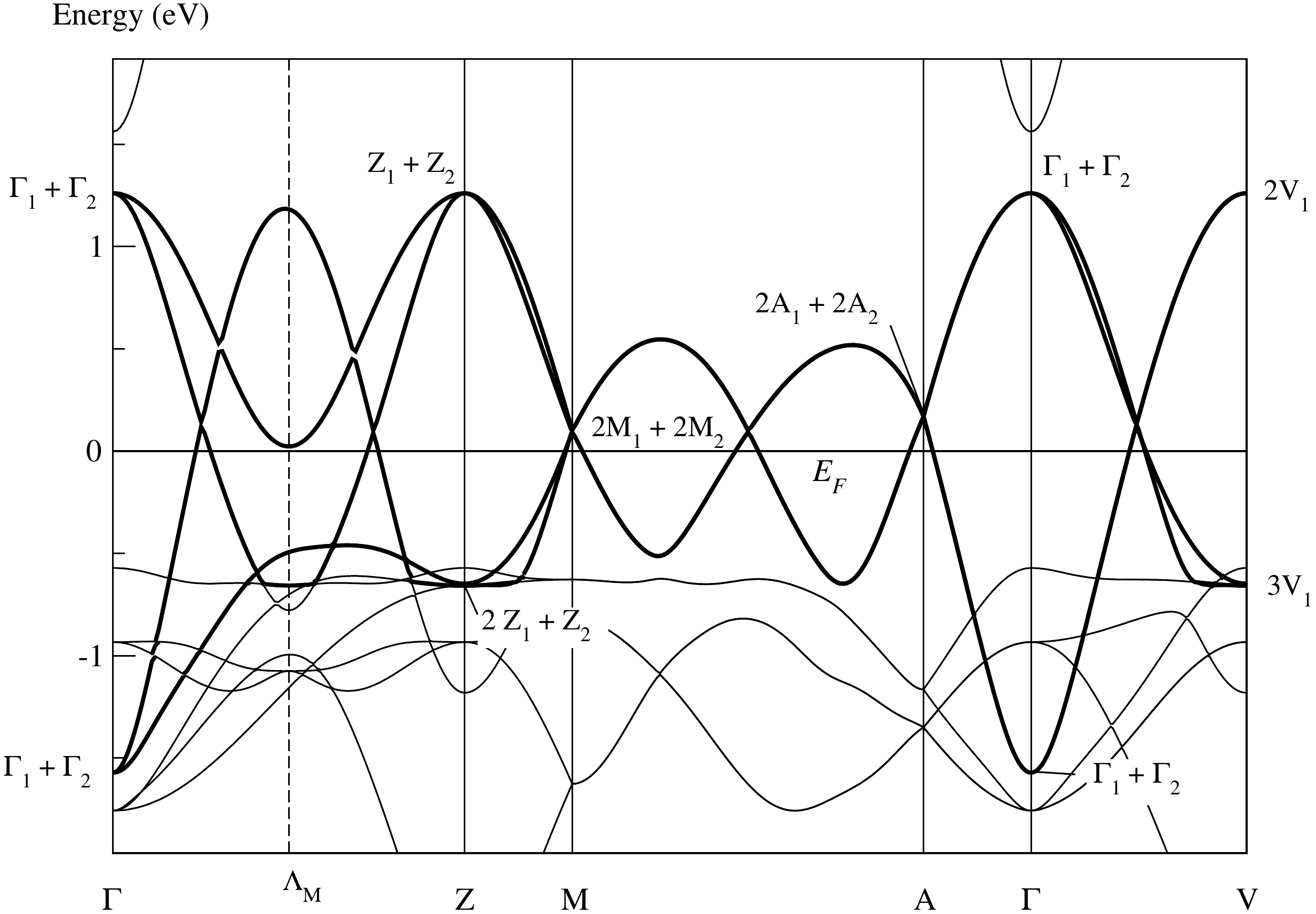}%
\caption{ The band structure of NiO given in Fig.\
 \ref{fig:bandstr225} folded into the
Brillouin zone for the monoclinic base centered  Bravais lattice $\Gamma^b_m$
of the magnetic group $M_9$. The band highlighted in
Fig.~\ref{fig:bandstr225} by the bold lines is
still highlighted by bold lines in the folded band structure. It now
forms a magnetic ``super'' band consisting of four branches assigned
to the two nickel and the two oxygen atoms.  
The symmetry labels are defined in
Table~\ref{tab:rep_9} and are determined from
Fig.~\ref{fig:bandstr225} by means of Table~\ref{tab:falten225_9}. The notations of
the points of symmetry follow Fig.~3.4~of Ref.~\cite{bc}. The
midpoint $\Lambda_{\text{M}}$ of the line $\overline{\Gamma Z}$ is equivalent
to the points $W' (\overline{\frac{1}{4}}\frac{1}{4}\frac{1}{2})$ and
$\Sigma' (\frac{1}{4}\overline{\frac{1}{4}}0)$ in the Brillouin zone
for the paramagnetic fcc lattice.  
\label{fig:bandstr9}
}
 \end{figure*}

All the information we now need is given in Fig.~\ref{fig:bandstr9}
and in Table~\ref{tab:wf_9}. When folding the band structure of
paramagnetic NiO given in Fig.~\ref{fig:bandstr225} into the Brillouin
zone of the monoclinic-base centered magnetic structure, we receive
the bands plotted in Fig.~\ref{fig:bandstr9}. The band highlighted in
the paramagnetic band structure by the bold lines is still highlighted
in Fig.~\ref{fig:bandstr9}. In what follows, it is referred to as the
``active band''. In the same manner as in Ref.~\cite{theoriewf} (see
Definition 2 {\em ibidem}) we call the active band a ``single band''
consisting of two (Fig.~\ref{fig:bandstr225}) or four
(Fig.~\ref{fig:bandstr9}) branches. The active band in
Fig.~\ref{fig:bandstr9} is a magnetic band related to the magnetic
group $M_9$ because the Bloch functions of two branches of this band
bear the symmetry labels of band 1 in Table~\ref{tab:wf_9} (a). Thus,
we can unitarily transform the Bloch functions of two branches of the
active band into optimally localized Wannier functions
symmetry-adapted to $M_9$ and centered at the Ni atoms. Thus, the
active band provides localized states allowing the electrons to
perform a nonadiabatic atomic-like motion stabilized by the
nonadiabatic condensation energy $\Delta E$. However, a prerequisite
is that a magnetic structure with the magnetic group $M_9$ is actually
realized in the crystal. So the electron system activates a
spin-dependent exchange mechanism producing the magnetic structure
with the magnetic group $M_{15}$. Such a mechanism is possible within
the nonadiabatic system, see Sec.~2 of Ref.~\cite{ybacuo6}. The group
$M_{15}$ is reduced to the group $M_9$ by the dislocations of the Ni
atoms.


In summary, the electrons produce the magnetic structure with the
magnetic group $M_9$, so that the symmetry of the crystal is modified
in such a way that the active band becomes a {\em closed}
band~\cite{theoriewf}. This allows the electrons at the Fermi level to
occupy an atomic-like state stabilized by the nonadiabatic
condensation energy $\Delta E$~\cite{enhm}.

\section{Mott insulator -- third condition of stability}
\label{sec:mottinsulator}
The active band of NiO does not only contain a magnetic band but has
in addition two remarkable features:
\begin{enumerate}
\item The magnetic band with the symmetry in Table~\ref{tab:wf_9} (a)
  occurs {\em twice} in the active band. Since band 1 in
  Table~\ref{tab:wf_9} (b) has the same symmetry, we can unitarily
  transform the Bloch functions of two branches of the active band
  into optimally localized Wannier functions centered at the Ni atoms,
  and the Bloch functions of the remaining two branches into optimally
  localized Wannier functions centered at the O atoms. Thus, the
  electrons perform an atomic-like motion with localized states
  situated at both the Ni {\em and} the O atoms.
\item {\em All the electrons} at the Fermi level take part in the
  atomic-like motion because the active band consists of all the
  branches crossing the Fermi level.   
\end{enumerate}
Such an active band I already found in the band structure of
BaMn$_2$As$_2$~\cite{bamn2as2} and called it ``magnetic super
band'':
\begin{Definition}
  The Bloch functions of a magnetic super band related to the magnetic
  group $M$ can be unitarily transformed into optimally localized
  Wannier functions symmetry-adapted to $M$ in such a way that the
  Wannier functions are not only centered the atoms bearing the
  magnetic moments, but are centered at all the atoms of the
  material. Moreover, each Bloch function at the Fermi level belongs
  to the super band.
\end{Definition}

  Thus, a magnetic super band produces a nonadiabatic atomic-like
  motion not only between the atoms bearing the magnetic moments, but
  between {\em all the} atoms of the material, whereby all the electrons
  at the Fermi level take part in the atomic-like motion.
  A magnetic super band contains just as many branches as there
  are atoms in the unit cell, and only branches belonging to the super
  band cross the Fermi level.

If the magnetic super band of NiO is half-filled, it produces not only the
magnetic structure, but may produce an atomic-like state with exactly
one electron on each atom. If such a state has the lowest Coulomb
repulsion energy, the crystal is an Mott insulator, because there are
no further electrons at the Fermi level which would be able to carry
an electrical current.

Because NiO is a prototype Mott
insulator~\cite{gavriliuk,mott_1949,austin}, the magnetic super band
of NiO is evidently half-filled and the Mott condition is evidently
satisfied in this band.  As mentioned, I already found a magnetic
super band in the band structure of BaMn$_2$As$_2$~\cite{bamn2as2}. In
fact, also BaMn$_2$As$_2$ is an band gap insulator, often referred to
as a small band-gap semicondoctor~\cite{an,singh2}. These observations
on NiO and BaMn$_2$As$_2$ suggest that the nonadiabatic atomic-like
motion in these materials involving all the atoms and all the
electrons at the Fermi level, is one cause for the insulating ground
state:
\begin{thm}
\label{mottinsulator}
Let be given a magnetic material with the magnetic space group $M$ that
possesses bands crossing the Fermi level in its band structure (i.e.,
the material should be metallic under band theory).
This material may be in fact a band gap or Mott insulator, if there
exists a half-filled narrow magnetic super band related to $M$ in its
band structure.
\end{thm}
In both NiO and BaMn$_2$As$_2$, the atomic-like motion breaks down
above the N\'eel temperature, so both materials become
metallic.

\section{Results}
The paper is concerned with three features of antiferromagnetic NiO,
where two of them are very special:
\begin{itemize}
\item The rhombohedral-like deformation of antiferromagnetic NiO,
\item the stability of the antiferromagnetic state, and
\item the insulating ground state.  
\end{itemize}
\subsection{The rhombohedral-like deformation}
The time-inversion symmetry of the electronic Hamiltonian requires
that the magnetic group of the antiferromagnetic state possesses
special irreducible co-representations (Condition 1 in
Sec.~\ref{sec:first_stab_cond}) so that the antiferromagnetic state is
stable. The maximal group $M_{15}$ leaving the magnetic structure of
NiO invariant, however, does not possess such suitable irreducible
co-representations. It is the subgroup $M_9$ of $M_{15}$ that has
co-representations allowing a stable magnetic structure.

Thus, the Ni atoms are shifted from their positions in the space group
$M_{15}$ as indicated in Fig.~\ref{fig:structures} (a) in order that
$M_9$ is realized, i.e., in order that $M_9$ but not $M_{15}$ is the
magnetic space group of the crystal.  This distortion creates a
deformation of the crystal closely resembling a rhombohedral
deformation but, clearly, cannot produce a rhombohedral space group
because the magnetic space group of NiO still is the monoclinic
base-centered group $M_9$.  The rhombohedral-like deformation may be
called an ``inner'' deformation of $M_9$.
\subsection{The stability of the antiferromagnetic state}
In Sec.~\ref{sec:mband} we have shown that NiO possesses a narrow,
roughly half-filled magnetic band related to the magnetic group $M_9$
in its band structure.  Thus, the electrons may perform a nonadiabatic
atomic-like motion with the localized states centered at the Ni
atoms. This atomic-like motion stabilizes the antiferromagnetic
structure of NiO.

\subsection{The insulating ground state}
Moreover, the magnetic band of NiO is a magnetic {\em super}
band. This means that the electrons even perform a nonadiabatic
atomic-like motion with the localized states centered at both the Ni
and the O atoms and that all the electrons at the Fermi level take
part in the atomic-like motion. This is an optimal precondition for
NiO to be a Mott insulator. As is well-known, NiO is indeed a
prototype Mott insulator.

\section{Discussion}
\label{sec:discussion}
The nonadiabatic Heisenberg model (NHM) claims that the hypothetical
nonadiabatic localized states defining the atomic-like motion
(Sec.~\ref{sec:atomiclike}) are {\em physically existent}. Thus, the
nonadiabatic atomic-like motion is {\em actually realized} at the
Fermi level if the considered metal possesses a narrow, partly filled
band with suitable exact Wannier functions
(Sec.~\ref{sec:atomiclike}). The success in the last 40 years in
identifying narrow, roughly half-filled
superconducting~\cite{ebi,theoriewf} and
magnetic~\cite{bamn2as2,theoriewf} bands in the band structures of
superconducting and magnetic materials provides evidence that the
nonadiabatic atomic-like motion as defined in the NHM actually has
physical reality and stabilizes the superconducting and magnetic
states, respectively. This statement is again corroborated by the
observation that NiO possesses a roughly half-filled magnetic band in
its band structure which is related to the magnetic group $M_9$ of the
antiferromagnetic structure. In addition, the NHM predicts that an
atomic-like motion exists in NiO (Sec.~\ref{sec:mottinsulator}) as
well as in BaMn$_2$As$_2$~\cite{bamn2as2} which involves {\em all the}
atoms and {\em all the} electrons at the Fermi level. Thus, the
electronic motion in these materials provides an ideal precondition for
the applicability of the Mott condition.  The observation that both
materials indeed are insulators may be understood as direct
experimental evidence for the atomic-like state in these materials.

The NHM provides the group-theoretical framework of any atomic-like
motion of the electrons at the Fermi level. It considers the {\em
  exact} atomic-like motion with hypothetical nonadiabatic localized
functions being not suited for the calculation of matrix elements. For
this purpose we still must approximately represent the nonadiabatic
localized states by atomic functions or (if they are known) by
adiabatic Wannier functions. This ``adiabatic approximation'' of
atomic-like electrons should yield physically relevant results if the
electrons {\em actually perform a nonadiabatic atomic-like motion}. On
the other hand, this adiabatic approximation should fail when the
electrons do not occupy an atomic-like state in the considered
material. Thus, the group-theoretical results presented in this paper
{\em create the basis} for any concept based on atomic-like electrons
in NiO. So, our results do not conflict with the concept of
correlation effects in narrow $d$ bands being responsible for the
nonmetallic behavior in NiO~\cite{austin,mott,mott_1961,mott_1949}.

\vspace{6pt} 




\acknowledgments{I am very indebted to Guido Schmitz for his support
  of my work.}

\conflictsofinterest{The author declares no conflict of interest.}
\abbreviations{The following abbreviations are used in this manuscript:\\

\noindent 
\begin{tabular}{@{}ll}
NHM & Nonadiabatic Heisenberg model\\
$E$ & Identity operation\\
$I$ & Inversion\\
$C_{2b}$ & Rotation through $\pi$ as indicated in Fig.~\ref{fig:structures}\\
$\sigma_{db}$ & Reflection $IC_{2b}$\\
$K$ & anti-unitary operator of time inversion
\end{tabular}}

\appendixtitles{no} 
\appendix
\section{Wyckoff positions}
\label{sec:wyckoff}
The magnetic group $M_9$ of the antiferromagnetic state of NiO is a
type III Shubnikov space group which also may be written in the
form~\cite{bc}
\begin{equation}
  \label{eq:21}
  M_9 = Cc + K(\overline{C2/c} - Cc),
\end{equation}
where the unitary space group $\overline{C2/c}$ contains (besides the
translations) the elements
\begin{equation}
   \label{eq:22}
 \overline{C2/c} = \Big\{\{E|\bm 0\}, \{C_{2b}|\bm 0\}, \{I|\bm \tau\},
 \{\sigma_{db}|\bm \tau\}\Big\}.
\end{equation}
Though the symmetry operations of the group $\overline{C2/c}$ are
different from the operations contained in $C2/c$ (see Eq.~\gl{eq:9}),
both groups bear the same international number 15 because the
elements of $\overline{C2/c}$ are changed into the elements of $C2/c$
when the origin of $\overline{C2/c}$ is translated by
\begin{equation}
  \label{eq:28}
  \bm t_0 = \frac{1}{4}\bm T_1
\end{equation}
~\cite{bc}. Keeping this in mind, we may
determine the Wyckoff positions of the atoms in the space group
$\overline{C2/c}$ by means of the Bilbao Crystallographic
Server~\cite{bilbao} yielding the Wyckoff positions $(\bm a|\bm b|\bm c)$ 
\begin{equation}
  \label{eq:27}
  \begin{array}{ccl}
  4b & (0 | \frac{1}{2}| 0) & (0 | \frac{1}{2}| \frac{1}{2})\\
  4e & (0 |y | \frac{1}{4}) & (0 | -y| \frac{3}{4})
  \end{array}
\end{equation}
in the space group $C2/c$. In the coordinate system given in 
Fig.~\ref{fig:structures} (a), the two Wyckoff positions $4b$
may be written as
\begin{equation}
  \label{eq:29}
  \begin{array}{cclcl}
  \bm p_{4b_1} & = & \frac{1}{2}(\bm T_2 - \bm T_3) + \bm t_0 & = & \frac{1}{2}(\bm T_2 - \bm T_3) + \frac{1}{4}\bm
  T_1\\
  \bm p_{4b_2} & = & \frac{1}{2}(\bm T_2 - \bm T_3) + \frac{1}{2}\bm
  T_1 + \bm t_0& = & \frac{1}{2}(\bm T_2 - \bm T_3) + \frac{1}{2}\bm
  T_1 + \frac{1}{4}\bm
  T_1,\\
  \end{array}
\end{equation}
and in the case $4e$ we have (for $y = 0$)  
\begin{equation}
  \label{eq:30}
  \begin{array}{cclcl}
  \bm p_{4e_1} & = & \frac{1}{4}\bm  T_1 + \bm t_0 & = & \frac{1}{2}\bm  T_1\\
  \bm p_{4e_2} & = & \frac{3}{4}\bm  T_1 + \bm t_0 & = & \bm
  T_1,
  \end{array}
\end{equation}
since (in Eq.~\gl{eq:27}) $\bm b = \bm T_2 - \bm T_3$ and  $\bm c = \bm T_1$.

Thus, the vectors $\bm t_{Ni}$ and $\bm t_{O}$ in Eqs.~\gl{eq:1}
and~\gl{eq:4} represent the Wyckoff positions $4e$ and $4b$ of
$\overline{C2/c}$. Eq.~\gl{eq:27} confirms that in the magnetic group
$M_9$ the nickel atoms may be shifted by $\pm y$ in
$\bm T_2 - \bm T_3$ direction from their positions $\bm t_{Ni}$, while
the oxygen atoms are fixed at their positions $\bm t_{O}$.
\section{Group-theoretical tables}
\label{sec:tables}
This appendix provides Tables~\ref{tab:rep_15} --~\ref{tab:wf_9}
along with notes to the tables.
\FloatBarrier

\begin{table}[h]
\caption{
  Character tables of the single-valued irreducible representations of the
  monoclinic base-centered space group $C2/c = \Gamma^b_mC^{6}_{2h}$ (15).
  \label{tab:rep_15}}
\begin{center}
\begin{tabular}[t]{cccccc}
\multicolumn{6}{c}{$\Gamma (000)$, $Z (0\overline{\frac{1}{2}}\frac{1}{2})$}\\
 & $K$ & $\{E|\bm 0\}$ & $\{C_{2b}|\bm\tau\}$ & $\{I|\bm 0\}$ & $\{\sigma_{db}|\bm \tau\}$\\
\hline
$\Gamma^+_1$, $Z^+_1$ & (a) & 1 & 1 & 1 & 1\\
$\Gamma^-_1$, $Z^-_1$ & (a) & 1 & 1 & -1 & -1\\
$\Gamma^+_2$, $Z^+_2$ & (a) & 1 & -1 & 1 & -1\\
$\Gamma^-_2$, $Z^-_2$ & (a) & 1 & -1 & -1 & 1\\
\hline\\
\end{tabular}\hspace{1cm}

\begin{tabular}[t]{ccccccc}
\multicolumn{7}{c}{$A (\overline{\frac{1}{2}}00)$, $M (\overline{\frac{1}{2}}\overline{\frac{1}{2}}\frac{1}{2})$}\\
 &  & $$ & $$ & $\{\sigma_{db}|3\bm \tau\}$ & $\{I|\bm T_1\}$ & $\{C_{2b}|3\bm \tau\}$\\
 & $K$ & $\{E|\bm 0\}$ & $\{E|\bm T_1\}$ & $\{\sigma_{db}|\bm \tau\}$ & $\{I|\bm 0\}$ & $\{C_{2b}|\bm 0\}$\\
\hline
$A_1$, $M_1$ & (a) & 2 & -2 & 0 & 0 & 0\\
\hline\\
\end{tabular}\hspace{1cm}
\end{center}
\ \\
\begin{flushleft}
Notes to Table~\ref{tab:rep_15}
\end{flushleft}
\begin{enumerate}
\item The notations of the points of symmetry follow Fig. 3.4 
of Ref.~\cite{bc}.
\item Only the points of symmetry invariant under the complete space
  group are listed.
\item The character tables are determined from Table 5.7 in
  Ref.~\protect\cite{bc}.  
\item $K$ denotes the operator of time inversion. The entry (a) is
  determined by equation\ (7.3.51) of Ref.~\cite{bc} and indicates
  that the related co-representations of the magnetic group
  $C2/c + KC2/c$ follow case (a) as defined in equation\ (7.3.45) of
  Ref.~\cite{bc}.
\end{enumerate}
\end{table}
\begin{table}[!]
\caption{
  Character tables of the single-valued irreducible representations of the
  monoclinic base-centered space group $Cc = \Gamma^b_mC^{4}_{1h}$ (9).
  \label{tab:rep_9}}
\begin{center}
\begin{tabular}[!]{cccccc}
\multicolumn{6}{c}{$\Gamma (000)$, $Z (0\overline{\frac{1}{2}}\frac{1}{2})$}\\
 & $K$ & $K\{E|\bm \tau\}$ & $K\{C_{2b}|\bm 0\}$ & $\{E|\bm 0\}$ & $\{\sigma_{db}|\bm \tau\}$\\
\hline
$\Gamma_1$, $Z_1$ & (a) & (a) & (a) & 1 & 1\\
$\Gamma_2$, $Z_2$ & (a) & (a) & (a) & 1 & -1\\
\hline\\
\end{tabular}\hspace{.4cm}
\begin{tabular}[!]{cc}
\multicolumn{2}{c}{$L (\overline{\frac{1}{2}}0\frac{1}{2})$, $V (00\frac{1}{2})$}\\
& $\{E|\bm 0\}$\\
\hline
$L_1$, $V_1$ &  1\\
\hline\\
\end{tabular}\hspace{1cm}
\begin{tabular}[!]{cccccccc}
\multicolumn{8}{c}{$A (\overline{\frac{1}{2}}00)$, $M (\overline{\frac{1}{2}}\overline{\frac{1}{2}}\frac{1}{2})$}\\
 & $K$ & $K\{E|\bm \tau\}$ & $K\{C_{2b}|\bm 0\}$ & $\{E|\bm 0\}$ & $\{\sigma_{db}|\bm \tau\}$ & $\{E|\bm T_1\}$ & $\{\sigma_{db}|3\bm \tau\}$\\
\hline
$A_1$, $M_1$ & (c) & (c) & (a) & 1 & i & -1 & -i\\
$A_2$, $M_2$ & (c) & (c) & (a) & 1 & -i & -1 & i\\
\hline\\
\end{tabular}\hspace{1cm}
\end{center}
\ \\
\begin{flushleft}
Notes to Table~\ref{tab:rep_9}
\end{flushleft}
\begin{enumerate}
\item The notations of the points of symmetry follow Fig. 3.4 
of Ref.~\cite{bc}.
\item The character tables are determined from Table 5.7 in
  Ref.~\protect\cite{bc}.  
\item $K$ denotes the operator of time inversion. The entries (a) and
  (c) are determined by equation\ (7.3.51) of Ref.~\cite{bc}. They
  indicate whether the related co-representations of the magnetic
  groups $Cc + KCc$, $Cc + K\{E|\bm \tau\}Cc$, and
  $Cc + K\{C_{2b}|\bm 0\}Cc$ follow case (a) or case (c) as defined in
  equations\ (7.3.45) and (7.3.47), respectively, of Ref.~\cite{bc}.
\end{enumerate}
\end{table}

\begin{table}[!]
\caption{
  Character tables of the single-valued irreducible representations of
  the cubic space group $Fm3m = \Gamma^f_cO^5_h$ (225) of 
  paramagnetic NiO. 
\label{tab:rep_225}}
\begin{tabular}[t]{ccccccccccc}
\multicolumn{11}{c}{$\Gamma (000)$}\\
 & $$ & $$ & $$ & $$ & $C^-_{34}$ & $S^+_{64}$ & $$ & $$ & $$ & $$\\
 & $$ & $$ & $$ & $$ & $C^+_{31}$ & $S^-_{61}$ & $$ & $$ & $$ & $$\\
 & $$ & $$ & $$ & $$ & $C^-_{32}$ & $S^+_{62}$ & $C^+_{4y}$ & $S^-_{4y}$ & $C_{2a}$ & $\sigma_{da}$\\
 & $$ & $$ & $$ & $$ & $C^+_{32}$ & $S^-_{62}$ & $C^-_{4y}$ & $S^+_{4y}$ & $C_{2f}$ & $\sigma_{df}$\\
 & $$ & $$ & $$ & $$ & $C^-_{33}$ & $S^+_{63}$ & $C^+_{4z}$ & $S^-_{4z}$ & $C_{2b}$ & $\sigma_{db}$\\
 & $$ & $$ & $\sigma_y$ & $C_{2y}$ & $C^+_{33}$ & $S^-_{63}$ & $C^+_{4x}$ & $S^-_{4x}$ & $C_{2d}$ & $\sigma_{dd}$\\
 & $$ & $$ & $\sigma_z$ & $C_{2z}$ & $C^-_{31}$ & $S^+_{61}$ & $C^-_{4z}$ & $S^+_{4z}$ & $C_{2e}$ & $\sigma_{de}$\\
 & $E$ & $I$ & $\sigma_x$ & $C_{2x}$ & $C^+_{34}$ & $S^-_{64}$ & $C^-_{4x}$ & $S^+_{4x}$ & $C_{2c}$ & $\sigma_{dc}$\\
\hline
$\Gamma^+_1$ & 1 & 1 & 1 & 1 & 1 & 1 & 1 & 1 & 1 & 1\\
$\Gamma^+_2$ & 1 & 1 & 1 & 1 & 1 & 1 & -1 & -1 & -1 & -1\\
$\Gamma^-_2$ & 1 & -1 & -1 & 1 & 1 & -1 & -1 & 1 & -1 & 1\\
$\Gamma^-_1$ & 1 & -1 & -1 & 1 & 1 & -1 & 1 & -1 & 1 & -1\\
$\Gamma^+_3$ & 2 & 2 & 2 & 2 & -1 & -1 & 0 & 0 & 0 & 0\\
$\Gamma^-_3$ & 2 & -2 & -2 & 2 & -1 & 1 & 0 & 0 & 0 & 0\\
$\Gamma^+_4$ & 3 & 3 & -1 & -1 & 0 & 0 & 1 & 1 & -1 & -1\\
$\Gamma^+_5$ & 3 & 3 & -1 & -1 & 0 & 0 & -1 & -1 & 1 & 1\\
$\Gamma^-_4$ & 3 & -3 & 1 & -1 & 0 & 0 & 1 & -1 & -1 & 1\\
$\Gamma^-_5$ & 3 & -3 & 1 & -1 & 0 & 0 & -1 & 1 & 1 & -1\\
\hline\\
\end{tabular}\hspace{1cm}
\begin{tabular}[t]{ccccccc}
\multicolumn{7}{c}{$L (\frac{1}{2}\frac{1}{2}\frac{1}{2})$}\\
 & $$ & $$ & $$ & $$ & $C_{2e}$ & $\sigma_{db}$\\
 & $$ & $$ & $S^-_{61}$ & $C^-_{31}$ & $C_{2f}$ & $\sigma_{de}$\\
 & $E$ & $I$ & $S^+_{61}$ & $C^+_{31}$ & $C_{2b}$ & $\sigma_{df}$\\
\hline
$L^+_1$ & 1 & 1 & 1 & 1 & 1 & 1\\
$L^+_2$ & 1 & 1 & 1 & 1 & -1 & -1\\
$L^-_1$ & 1 & -1 & -1 & 1 & 1 & -1\\
$L^-_2$ & 1 & -1 & -1 & 1 & -1 & 1\\
$L^+_3$ & 2 & 2 & -1 & -1 & 0 & 0\\
$L^-_3$ & 2 & -2 & 1 & -1 & 0 & 0\\
\hline\\
\end{tabular}\hspace{1cm}
\begin{tabular}[t]{ccccccccccc}
\multicolumn{11}{c}{$X (\frac{1}{2}0\frac{1}{2})$}\\
 & $$ & $$ & $C^-_{4y}$ & $C_{2z}$ & $C_{2c}$ & $$ & $$ & $S^+_{4y}$ & $\sigma_z$ & $\sigma_{dc}$\\
 & $E$ & $C_{2y}$ & $C^+_{4y}$ & $C_{2x}$ & $C_{2e}$ & $I$ & $\sigma_y$ & $S^-_{4y}$ & $\sigma_x$ & $\sigma_{de}$\\
\hline
$X^+_1$ & 1 & 1 & 1 & 1 & 1 & 1 & 1 & 1 & 1 & 1\\
$X^+_2$ & 1 & 1 & 1 & -1 & -1 & 1 & 1 & 1 & -1 & -1\\
$X^+_3$ & 1 & 1 & -1 & 1 & -1 & 1 & 1 & -1 & 1 & -1\\
$X^+_4$ & 1 & 1 & -1 & -1 & 1 & 1 & 1 & -1 & -1 & 1\\
$X^+_5$ & 2 & -2 & 0 & 0 & 0 & 2 & -2 & 0 & 0 & 0\\
$X^-_1$ & 1 & 1 & 1 & 1 & 1 & -1 & -1 & -1 & -1 & -1\\
$X^-_2$ & 1 & 1 & 1 & -1 & -1 & -1 & -1 & -1 & 1 & 1\\
$X^-_3$ & 1 & 1 & -1 & 1 & -1 & -1 & -1 & 1 & -1 & 1\\
$X^-_4$ & 1 & 1 & -1 & -1 & 1 & -1 & -1 & 1 & 1 & -1\\
$X^-_5$ & 2 & -2 & 0 & 0 & 0 & -2 & 2 & 0 & 0 & 0\\
\hline\\
\end{tabular}\hspace{1cm}
\begin{tabular}[t]{cccccc}
\multicolumn{6}{c}{$W (\frac{1}{2}\frac{1}{4}\frac{3}{4})$}\\
 & $$ & $$ & $S^-_{4x}$ & $C_{2f}$ & $\sigma_z$\\
 & $E$ & $C_{2x}$ & $S^+_{4x}$ & $C_{2d}$ & $\sigma_y$\\
\hline
$W_1$ & 1 & 1 & 1 & 1 & 1\\
$W_2$ & 1 & 1 & 1 & -1 & -1\\
$W_3$ & 1 & 1 & -1 & 1 & -1\\
$W_4$ & 1 & 1 & -1 & -1 & 1\\
$W_5$ & 2 & -2 & 0 & 0 & 0\\
\hline\\
\end{tabular}\hspace{1cm}
\begin{tabular}[t]{cccc}
\multicolumn{4}{c}{$\Lambda (\frac{1}{4}\frac{1}{4}\frac{1}{4})$}\\
 & $$ & $$ & $\sigma_{df}$\\
 & $$ & $C^-_{31}$ & $\sigma_{de}$\\
 & $E$ & $C^+_{31}$ & $\sigma_{db}$\\
\hline
$\Lambda_1$ & 1 & 1 & 1\\
$\Lambda_2$ & 1 & 1 & -1\\
$\Lambda_3$ & 2 & -1 & 0\\
\hline\\
\end{tabular}\hspace{1cm}
\begin{tabular}[t]{ccc}
\multicolumn{3}{c}{$R (\frac{1}{4}\frac{1}{4}\frac{3}{4})$}\\
 & $E$ & $\sigma_{db}$\\
\hline
$R_1$ & 1 & 1\\
$R_2$ & 1 & -1\\
\hline\\
\end{tabular}\hspace{1cm}
\ \\
\begin{flushleft}
Notes to Table~\ref{tab:rep_225}
\end{flushleft}
\begin{enumerate}
\item The point group operations are related to the $x, y, z$
  coordinate system in Fig.~\ref{fig:structures}.
\item The notations of the points of symmetry follow Fig. 3.14 of 
Ref.~\cite{bc}.
\item The character tables are determined from Table 5.7 of
  Ref.~\protect\cite{bc}.
\item The point $R$ lies in the plane $\Gamma L K$ 
\end{enumerate}
\end{table}

\begin{table}[!]
\caption{
Compatibility relations between the Brillouin zone for the fcc space group
$Fm3m$ (225) of paramagnetic NiO and the Brillouin zone for the space group $Cc$ (9) of the
  antiferromagnetic
  structure in distorted NiO.
\label{tab:falten225_9}
}
\begin{flushleft}
\begin{tabular}[t]{cccccccccc}
\multicolumn{10}{c}{$\Gamma (000)$}\\
\hline
$\Gamma^+_1$ & $\Gamma^+_2$ & $\Gamma^-_2$ & $\Gamma^-_1$ & $\Gamma^+_3$ & $\Gamma^-_3$ & $\Gamma^+_4$ & $\Gamma^+_5$ & $\Gamma^-_4$ & $\Gamma^-_5$\\
$\Gamma_1$ & $\Gamma_2$ & $\Gamma_1$ & $\Gamma_2$ & $\Gamma_1$ + $\Gamma_2$ & $\Gamma_1$ + $\Gamma_2$ & $\Gamma_1$ + 2$\Gamma_2$ & 2$\Gamma_1$ + $\Gamma_2$ & 2$\Gamma_1$ + $\Gamma_2$ & $\Gamma_1$ + 2$\Gamma_2$\\
\hline\\
\end{tabular}\hspace{1cm}
\begin{tabular}[t]{cccccc}
\multicolumn{6}{c}{$L (\frac{1}{2}\frac{1}{2}\frac{1}{2})$}\\
\hline
$L^+_1$ & $L^+_2$ & $L^-_1$ & $L^-_2$ & $L^+_3$ & $L^-_3$\\
$\Gamma_2$ & $\Gamma_1$ & $\Gamma_1$ & $\Gamma_2$ & $\Gamma_1$ + $\Gamma_2$ & $\Gamma_1$ + $\Gamma_2$\\
\hline\\
\end{tabular}\hspace{1cm}
\begin{tabular}[t]{cccccc}
\multicolumn{6}{c}{$L' (00\frac{1}{2})$}\\
\hline
$L^+_1$ & $L^+_2$ & $L^-_1$ & $L^-_2$ & $L^+_3$ & $L^-_3$\\
$Z_2$ & $Z_1$ & $Z_1$ & $Z_2$ & $Z_1$ + $Z_2$ & $Z_1$ + $Z_2$\\
\hline\\
\end{tabular}\hspace{1cm}
\begin{tabular}[t]{cccccc}
\multicolumn{6}{c}{$L'' (\frac{1}{2}00)$}\\
\hline
$L^+_1$ & $L^+_2$ & $L^-_1$ & $L^-_2$ & $L^+_3$ & $L^-_3$\\
$V_1$ & $V_1$ & $V_1$ & $V_1$ & 2$V_1$ & 2$V_1$\\
\hline\\
\end{tabular}\hspace{1cm}
\begin{tabular}[t]{ccc}
\multicolumn{3}{c}{$\Lambda (\frac{1}{4}\frac{1}{4}\frac{1}{4})$}\\
\hline
$\Lambda_1$ & $\Lambda_2$ & $\Lambda_3$\\
$A_1$ + $A_2$ & $A_1$ + $A_2$ & 2$A_1$ + 2$A_2$\\
\hline\\
\end{tabular}\hspace{1cm}
\begin{tabular}[t]{cc}
\multicolumn{2}{c}{$R (\frac{1}{4}\frac{1}{4}\frac{3}{4})$}\\
\hline
$R_1$ & $R_2$\\
$M_1$ + $M_2$ & $M_1$ + $M_2$\\
\hline\\
\end{tabular}\hspace{1cm}
\begin{tabular}[t]{cccccccccc}
\multicolumn{10}{c}{$X' (\overline{\frac{1}{2}}\overline{\frac{1}{2}}0)$}\\
\hline
$X^+_1$ & $X^+_2$ & $X^+_3$ & $X^+_4$ & $X^+_5$ & $X^-_1$ & $X^-_2$ & $X^-_3$ & $X^-_4$ & $X^-_5$\\
$Z_1$ & $Z_2$ & $Z_2$ & $Z_1$ & $Z_1$ + $Z_2$ & $Z_2$ & $Z_1$ & $Z_1$ & $Z_2$ & $Z_1$ + $Z_2$\\
\hline\\
\end{tabular}\hspace{1cm}
\begin{tabular}[t]{cccccccccc}
\multicolumn{10}{c}{$X'' (0\overline{\frac{1}{2}}\overline{\frac{1}{2}})$}\\
\hline
$X^+_1$ & $X^+_2$ & $X^+_3$ & $X^+_4$ & $X^+_5$ & $X^-_1$ & $X^-_2$ & $X^-_3$ & $X^-_4$ & $X^-_5$\\
$V_1$ & $V_1$ & $V_1$ & $V_1$ & 2$V_1$ & $V_1$ & $V_1$ & $V_1$ & $V_1$ & 2$V_1$\\
\hline\\
\end{tabular}\hspace{1cm}
\end{flushleft}
\ \\
\begin{flushleft}
Notes to Table~\ref{tab:falten225_9}
\end{flushleft}
\begin{enumerate}
\item The Brillouin zone for $Cc$ (9) lies diagonally within the
  Brillouin zone for $Fm3m$ (225).
\item The upper rows list the representations of the little groups of the
  points of symmetry in the Brillouin zone for $Fm3m$ and the lower rows list
  representations of the little groups of the related points of symmetry in
  the Brillouin zone for $Cc$.
  
  The representations in the same column are compatible in the
  following sense: Bloch functions that are basis functions of a
  representation $\bm{D}_i$ in the upper row can be unitarily transformed into
  the basis functions of the representation given below $\bm{D}_i$.
\item The notations of the points of symmetry follow Fig. 3.14 and
  Fig. 3.4, respectively, of Ref.~\cite{bc}.
\item The notations of the representations are defined in
  Tables~\ref{tab:rep_225} and~\ref{tab:rep_9}, respectively.
\item Within the Brillouin zone for $Fm3m$ the primed points are
  equivalent to the unprimed point.  
\item The compatibility relations are determined by a C++ computer
  program in the way described in great detail in Ref.~\cite{eabf}.
\end{enumerate}
\end{table}

\begin{table}[!]
\caption{
Symmetry labels of the Bloch functions at the points of symmetry in
the Brillouin zone for $Cc$ (9) of all
the energy bands with symmetry-adapted and optimally  
localized Wannier functions centered at the Ni (Table (a)) and O
(Table (b)) atoms, respectively. 
\label{tab:wf_9}}
\begin{flushleft}
\begin{tabular}[t]{cccccccccc}
(a)\ \ {\bf Ni} & Ni$_1(000)$ & Ni$_2(\overline{\frac{1}{2}}00)$ & $K\{C_{2b}|\bm
0\}$ & $\Gamma$ & $A$ & $Z$ & $M$ & $L$  & $V$\\
\hline
Band 1 & $\bm{d}_{1}$ & $\bm{d}_{1}$ & OK & $\Gamma_1$ + $\Gamma_2$ & $A_1$ + $A_2$ & $Z_1$ + $Z_2$ & $M_1$ + $M_2$ & 2$L_1$ & 2$V_1$\\
\hline\\
\end{tabular}\hspace{1cm}
\begin{tabular}[t]{cccccccccc}
(b)\ \ {\bf O} & O$_1(\overline{\frac{1}{4}}\frac{1}{2}\overline{\frac{1}{2}})$ & O$_2(\overline{\frac{3}{4}}\frac{1}{2}\overline{\frac{1}{2}})$ & $K\{C_{2b}|\bm
0\}$ & $\Gamma$ & $A$ & $Z$ & $M$ & $L$ & $V$\\
\hline
Band 1 & $\bm{d}_{1}$ & $\bm{d}_{1}$ &  OK  & $\Gamma_1$ + $\Gamma_2$ & $A_1$ + $A_2$ & $Z_1$ + $Z_2$ & $M_1$ + $M_2$ & 2$L_1$ & 2$V_1$\\
\hline\\
\end{tabular}\hspace{1cm}
\end{flushleft}
\begin{flushleft}
Notes to Table~\ref{tab:wf_9}
\end{flushleft}
\begin{enumerate}
\item The space group $Cc$ is the unitary subgroup of the magnetic
  group $M_9 = Cc + K\{C_{2b}|\bm 0\}Cc$ leaving invariant both the experimentally
  observed \cite{rothI,rothII,yamadaI,yamadaII,cj} antiferromagnetic
  structure and the dislocations of the Ni atoms shown in
  Fig.~\ref{fig:structures} (a). $K$ still denotes
  the operator of time-inversion.
\item Each band consists of two branches (Definition 2 of Ref.\
  \cite{theoriewf}) since there are two Ni and two O atoms in the unit
  cell.
\item Band 1 of Ni forms the magnetic band responsible for the
  antiferromagnetic structure of NiO.
\item Band 1 of Ni and band 1 of O form together the magnetic super
  band responsible for the Mott insulator.
\item The notations of the points of symmetry in the Brillouin zone
  for $\Gamma^b_m$ follow Fig. 3.4 of Ref.~\cite{bc}.
\item The symmetry notations of the Bloch functions are defined in
  Table~\ref{tab:rep_9}.
\item The bands are determined following Theorem 5 of Ref.\
  \cite{theoriewf}.
\item Table (a) is valid irrespective of whether or not the Ni atoms are
  dislocated as shown in Fig.~\ref{fig:structures} (a).  
\item The Wannier functions at the Ni or O atoms listed in the
  upper row belong to the representation $\bm{d}_1$ included below the atom. 
\item Applying Theorem 5, we need the representation $\bm d_1$ of the
  point groups $G_{0Ni}$ and $G_{0O}$ of the positions of the Ni and O
  atoms, respectively (Definition 12 of Ref.\ \cite{theoriewf}). In
  NiO, both groups contain only the identity operation,
  \begin{equation}
    \label{eq:24}
    G_{0Ni} = G_{0O} = \Big\{ \{E|\bm 0\} \Big\}.
    \end{equation}
    Thus, the Wannier functions belong to the simple representation
    defined by
\begin{center}
\begin{tabular}[t]{cc}
 & $\{E|\bm 0\}$\\
\hline
$\bm{d}_{1}$ & 1\\
\hline\\
\end{tabular}\hspace{1cm}
\end{center}
\item Each row defines a band with Bloch functions that can be
  unitarily transformed into Wannier functions being
\begin{itemize}
\item as well localized as possible (according to Definition 5 of Ref.~\cite{theoriewf}); 
\item centered at the Ni (Table (a)) or O (Table (b)) atoms; and
\item symmetry-adapted to $Cc$. That means (Definition 7 of
  Ref.~\cite{theoriewf}) that they satisfy Equation (15) of
  Ref.~\cite{theoriewf}, reading in NiO as
  \begin{equation}
    \label{eq:25}
    \begin{array}{lcl}
    P(\{\sigma_{db}|\bm \tau\})w_{Ni_1}(\bm r) & = & w_{Ni_2}(\bm r), \\
    P(\{\sigma_{db}|\bm \tau\})w_{Ni_2}(\bm r) & = & w_{Ni_1}(\bm r), \\
    P(\{\sigma_{db}|\bm \tau\})w_{O_1}(\bm r) & = & w_{O_2}(\bm r),  \\ 
    P(\{\sigma_{db}|\bm \tau\})w_{O_2}(\bm r) & = & w_{O_1}(\bm r),
    \end{array}
  \end{equation}
  where $w_{Ni_1}(\bm r), w_{Ni_2}(\bm r), w_{O_1}(\bm r)$, and
  $w_{O_2}(\bm r)$ denote the Wannier functions centered at the Ni and
  O atoms, respectively. 
\end{itemize}
\end{enumerate}
\end{table}

\begin{table}[t]
\begin{flushleft}
Notes to Table~\ref{tab:wf_9} (continued)
\end{flushleft}
\begin{enumerate}
\setcounter{enumi}{11}
\item The entry ``OK'' indicates that the Wannier functions follow
  also Theorem 7 of Ref.\ \cite{theoriewf} with
  $\bm N = \Big(\begin{array}{cc}1 & 0\\0 & 1\end{array}\Big)$ in
  Table (a) and
  $\bm N = \Big(\begin{array}{cc}0 & 1\\1 & 0\end{array}\Big)$ in
  Table (b). That means that the Wannier functions may even be chosen
  symmetry-adapted to the magnetic group
  $M = Cc + K\{C_{2b}|\bm 0\}Cc$. Thus, Eq. (62) of
  Ref.~\cite{theoriewf} is valid, reading in NiO as
  \begin{equation}
    \label{eq:26}
    \begin{array}{lcl}
    KP(\{C_{2b}|\bm 0\})w_{Ni_{1}}(\bm r) & = & w_{Ni_{1}}(\bm r)\\
    KP(\{C_{2b}|\bm 0\})w_{Ni_{2}}(\bm r) & = & w_{Ni_{2}}(\bm r)\\
    KP(\{C_{2b}|\bm 0\})w_{O_{1}}(\bm r) & = & w_{O_{2}}(\bm r) \\
    KP(\{C_{2b}|\bm 0\})w_{O_{2}}(\bm r) & = & w_{O_{1}}(\bm r)
    \end{array}
  \end{equation}
in addition to Eqs.~\gl{eq:25}. 
\item Within the NHM, the Eqs.~\gl{eq:25} and~\gl{eq:26} have only
  one, but important meaning: they ensure that the nonadiabatic
  Hamiltonian of the atomic-like electrons commutes with the symmetry
  operators of $M_9$~\cite{enhm}.
\end{enumerate}
\end{table}
\FloatBarrier

\bibliographystyle{mdpi}
\reftitle{References}


\externalbibliography{yes}





\end{document}